\documentclass[conference]{IEEEtran}
\IEEEoverridecommandlockouts

\usepackage{amsmath,amssymb,amsfonts}
\usepackage{graphicx}
\usepackage{stmaryrd}
\usepackage{textcomp}
\usepackage{xcolor}
\usepackage{comment}
\usepackage[none]{hyphenat}

\def\BibTeX{{\rm B\kern-.05em{\sc i\kern-.025em b}\kern-.08em
    T\kern-.1667em\lower.7ex\hbox{E}\kern-.125emX}}

\usepackage{array}
\usepackage{tabularx}
\usepackage{multirow}
\usepackage{booktabs}
\usepackage{colortbl}

\definecolor{gold}{RGB}{221, 196, 65}

\begin{document}

\title{Feasibility Analysis of Federated Neural Networks for Explainable Detection of Atrial Fibrillation}

\author{
\IEEEauthorblockN{
Diogo Reis Santos\IEEEauthorrefmark{1}, 
Andrea Protani\IEEEauthorrefmark{1}\IEEEauthorrefmark{2}, 
Lorenzo Giusti\IEEEauthorrefmark{1}, 
Albert Sund Aillet\IEEEauthorrefmark{1} \\
Pierpaolo Brutti\IEEEauthorrefmark{2}, 
Luigi Serio\IEEEauthorrefmark{1}} \\
\IEEEauthorblockA{\IEEEauthorrefmark{1}CERN, Geneva, Switzerland}
\IEEEauthorblockA{\IEEEauthorrefmark{2}Sapienza University of Rome, Rome, Italy}
}

\maketitle

\begin{abstract}
Early detection of atrial fibrillation (AFib) is challenging due to its asymptomatic and paroxysmal nature. However, advances in deep learning algorithms and the vast collection of electrocardiogram (ECG) data from devices such as the Internet of Things (IoT) hold great potential for the development of an effective solution. This study assesses the feasibility of training a neural network on a Federated Learning (FL) platform to detect AFib using raw ECG data. The performance of an advanced neural network is evaluated in centralized, local, and federated settings. The effects of different aggregation methods on model performance are investigated, and various normalization strategies are explored to address issues related to neural network federation. The results demonstrate that federated learning can significantly improve the accuracy of detection over local training. The best performing federated model achieved an F1 score of 77\%, improving performance by 15\% compared to the average performance of individually trained clients. This study emphasizes the promise of FL in medical diagnostics, offering a privacy-preserving and interpretable solution for large-scale healthcare applications.
\end{abstract}


\section{Introduction}

Rapid advances in machine learning (ML) offer great potential for biomedical data analysis and improving healthcare. ML solutions have the ability to improve patient care by processing and analyzing complex biomedical data, providing diagnostic support, and offering personalized treatment recommendations to improve patient outcomes and reduce costs~\cite{Shaheen2021ApplicationsReview, AlKuwaiti2023AHealthcare}. Atrial fibrillation (AFib) is the most common cardiac arrhythmia, affecting more than 33 million people worldwide. In the USA alone, it is projected to affect more than 5 million people by 2025 due to an aging population~\cite{Gillis2013ManagementMedicine}. This condition is associated with various comorbidities, cardiovascular complications, increased mortality, and, most significantly, a higher risk of stroke~\cite{Yiin2014Age-specificStudy}. Therefore, improving cardiac monitoring for early detection of AFib is crucial to mitigate these risks and improve patient outcomes~\cite{Attia2019AnPrediction}. However, early detection of AFib is challenging due to its asymptomatic and paroxysmal nature.

Traditional machine learning methods for detecting AFib, such as random forests and support vector machines, rely heavily on feature extraction processes from electrocardiogram (ECG) signals by employing techniques such as wavelets, Fourier transforms, and point detection~\cite{AlKuwaiti2023AHealthcare}. These methods often struggle to be generalized to real-world scenarios due to their tendency to overfit to limited and carefully curated datasets that may not accurately represent real-world data. In recent years, the application of deep learning techniques, particularly convolutional neural networks (CNN), in the context of AFib detection has shown promising performance by eliminating the need for complex feature extraction stages~\cite{Wegner2022MachineFibrillation}. However, training and validation of these models require access to extensive and diverse datasets, raising substantial concerns regarding privacy and security.

Federated Learning (FL) addresses these challenges by enabling the training of AI models across distributed nodes without the need to transfer local data~\cite{Rieke2020TheLearning}. Initially designed for mobile devices, FL has gained prominence in the medical field due to its privacy-preserving capabilities~\cite{Rieke2020TheLearning}. Recent research has shown that FL can achieve performance comparable to centralized models while preserving data privacy, making it an ideal approach to handling sensitive medical data~\cite{StoklasaBRAINLEARNING}. This study explores the application of FL in the detection of AFib from raw ECG data, leveraging the privacy-preserving and collaborative nature of FL to enhance the robustness and generalizability of the model.

The increasing popularity of devices on the Internet of Things (IoT), such as smartwatches and other wearable technologies, has led to a more standardized and widespread collection of ECG data~\cite{perez2019large, Guo2019MobileFibrillation}. These devices allow for continuous monitoring, which enables the gathering of large datasets needed for reliable AFib detection models. The standardized collection of ECG data from IoT devices also facilitates the use of FL by creating cohesive but distributed training datasets.

\section{Related works}

Recent studies in digital healthcare have shown that machine learning techniques can accurately identify cardiac abnormalities in ECG, achieving performance comparable to that of clinical cardiologists~\cite{hannun2019cardiologist}. The 2017 PhysioNet Computing in Cardiology Challenge (CINC) reported models with a maximum F1 score of 83.1\%~\cite{CliffordAF2017}. Techniques range from manually engineered features with random forest or gradient-boosting models to convolutional and recurrent neural networks. Despite good performance, a significant drop from validation to test sets highlighted the need for a comprehensive evaluation to prevent overfitting. Subsequent ensemble methods improved performance to around 90\%. The dataset continues to aid in the development and validation of algorithms~\cite{Isaksen2022}. Large-scale studies by Apple and Huawei have also been conducted. The Apple Heart study involved 419,000 participants using Apple smartwatches, detecting AFib in 34\% of notified cases~\cite{perez2019large}. The Huawei Heart study with 190,000 participants using Huawei smartwatches found AFib in nearly 90\% of notified cases~\cite{Guo2019MobileFibrillation}. However, these datasets are not publicly available, limiting external development and validation. The success of these studies underscores the potential of IoT devices, particularly smartwatches, for early detection of AFib. The issues of data availability, privacy, and ownership could be addressed with robust and secure federated learning platforms.

\section{Methods}

\subsection{Dataset}

\begin{figure}[t]
    \centering
    \includegraphics[width=\linewidth]{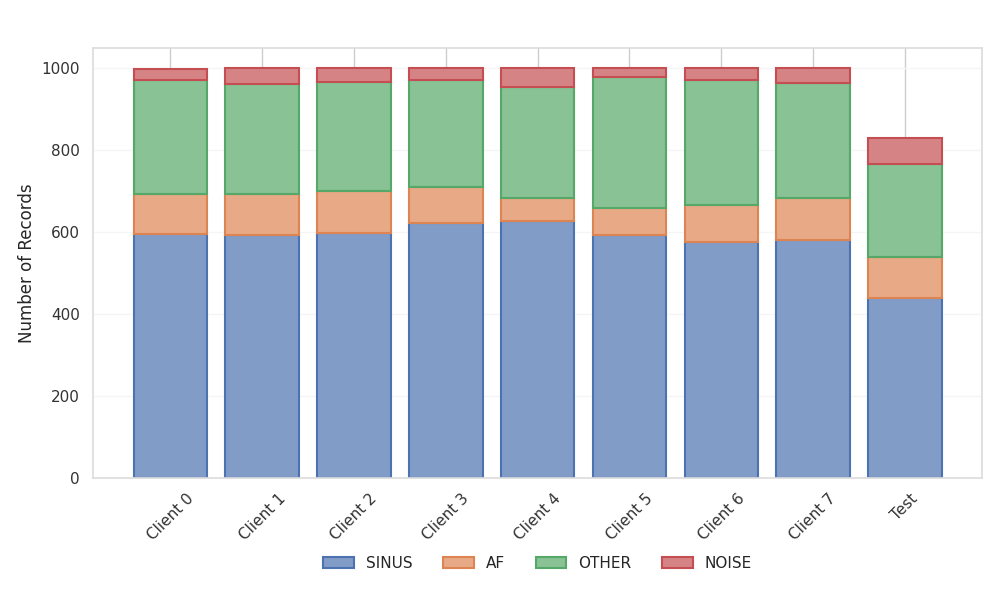}
    \vspace{-15pt}
     \caption{The distribution of ECG records by class (SINUS, AFib, OTHER, NOISE) for each client and the test set.}
     \label{fig:distr}
     \vspace{-10pt}
\end{figure}

We leveraged the publicly available data from the 2017 PhysioNet Challenge~\cite{CliffordAF2017}, consisting of ECG recordings from AliveCor devices. The original training set includes 8,528 single-lead ECG recordings, while the validation set contains 300 recordings, both lasting 9 to 60 seconds. The training set was divided into 9 parts, the first 8 containing 1,000 signals each and the last part containing 530. We merged the last part and the validation set to create a larger test set, addressing the discrepancy between evaluation and test performance. The first eight parts were used to simulate eight virtual data nodes, helping to facilitate future reproducibility studies as shown in Figure~\ref{fig:distr}. ECG recordings were originally sampled at 300 Hz and bandpass filtered by the AliveCor device. Labels were categorized as: (a) normal sinus rhythm (SINUS), (b) atrial fibrillation (AFib), (c) other rhythm (OTHER), or (d) too noisy to classify (NOISE). The data was resampled to 200 Hz to match the frequency used to develop the original ML model architecture. Approximately 75\% of the signals are 30 seconds long, with shorter signals padded with zeros and longer signals truncated to the central 30 seconds. No additional pre-processing was performed and raw ECG data was used as the model input.

\subsection{Model Architecture}

We have employed a CNN architecture designed for the detection of cardiac arrhythmias in ECG signals~\cite{hannun2019cardiologist}. The original architecture of this network takes raw ECG signal time series as input and generates a sequence of label predictions at 1.28-second intervals. However, in our dataset, each signal is associated with a single label. Therefore, instead of producing a sequence of predictions, we opted to flatten the output of the final convolutional block to obtain a single prediction for each signal. The CNN comprises 33 convolutional layers organized into 16 residual blocks. All these layers have a filter length of 16 and 64 k filters, where k starts from 1 and increases every fourth residual block. Each alternate residual block subsamples its input by a factor of 2, with the original input undergoing a total subsampling factor of $2^8$. After observing local overfitting in certain scenarios, we decided to reduce the model size by subsampling every block instead of every other block and reducing the total number of blocks by half. This \textit{'small network'} adjustment effectively reduces the number of parameters in the network by almost half while maintaining the output size.

\subsection{Federated Learning Platform} 

Our experiments utilize the CAFEIN FL platform~\cite{cafein}, developed at CERN, specifically designed to facilitate secure, robust, and privacy-preserving FL processes. The FL platform makes use of the Message Queuing Telemetry Transport (MQTT) protocol to exchange parameters between the clients and the server. Its support for the publish-subscribe messaging pattern is ideal for FL scenarios, allowing efficient and scalable communication between the parameter server and multiple client nodes. In our federated learning setup, every participating medical institution, known as a client node, maintains a local model that is trained on its private dataset. This distributed approach guarantees that sensitive patient data never leave the node premises. Instead of sharing raw data, the client nodes periodically send model updates (i.e. weights and gradients) to the central parameter server. The central parameter server aggregates these local model updates using aggregation algorithms.

\subsection{Experiments}

We evaluated the performance by examining three distinct scenarios: \textbf{local training}, distributed nodes utilize their local datasets to train local models without federated learning; \textbf{centralized training}, model training is performed on the entire centralized dataset. \textbf{federated training:} distributed nodes participate as federation members and implement a specified FL algorithm.

A federated system featuring a parameter server and $8$ virtual client nodes was implemented using Docker containers within a virtual machine, emulating a real-world federation environment. Architecture's weights are initialized using a Kaiming uniform strategy. The settings for the local, centralized, and federated experiments were aligned as closely as possible to ensure a fair comparison. Training was carried out for up to $256$ epochs or rounds. A learning rate reducer was set to activate after a plateau of $16$ epochs or rounds, and early stopping was configured for $48$ epochs or rounds. For federated aggregation, the four mentioned methods were evaluated. Batch normalization layers present inherent challenges in federated models, particularly with non-IDD data. Consequently, we assessed the model's performance under federated learning by replacing the batch normalization layers with group normalization and layer normalization \cite{casella2023experimenting}. Initially, Adam was used as the optimizer, as in the original network training \cite{hannun2019cardiologist}. However, most federated learning experiments employ SGD as the optimizer, since the training is paused and restarted at each round, resetting the gradients and momentum of Adam. We evaluated SGD with two distinct learning rates: the default $0.01$ and a reduced $0.001$.

\section{Results}
We evaluated the performance of a federated model using four different aggregation algorithms. The results, shown in Table~\ref{tab:results1}, are expressed as F1-score and accuracy. SCAFFOLD achieved the highest performance with an F1-score of 74\%, followed by FedAvg at 71\%. FedDyn and FedProx exhibited lower performance, resulting in their exclusion from further experiments.
Since BN layers are known to cause issues in federated settings, we evaluated the performance of layer and group normalization. Performance metrics are presented in Table~\ref{tab:results1}. Layer and group normalization led to lower performance compared to BN, with F1-scores dropping to 53\% and 37\%, respectively. Therefore, we utilized BN for subsequent experiments.
By default, we used Adam as the optimizer, but also evaluated SGD and SGD with a reduced learning rate. Both alternative optimizer configurations also resulted in lower performance, with F1-scores of 37\%. Table~\ref{tab:results2} summarizes the results obtained.

We observed significant overfitting to the training data of local nodes in most federated configurations, with metrics in local training reaching perfect values while evaluation metrics did not improve. Therefore, we decided to reduce the network size to the \textit{'small network'}. The results are shown in Table \ref{tab:results3}. For FedAvg, the performance decreased, but for SCAFFOLD, the performance improved from 74\% to 77\%.

\begin{figure}[t]
    \centering
    \includegraphics[width=\linewidth]{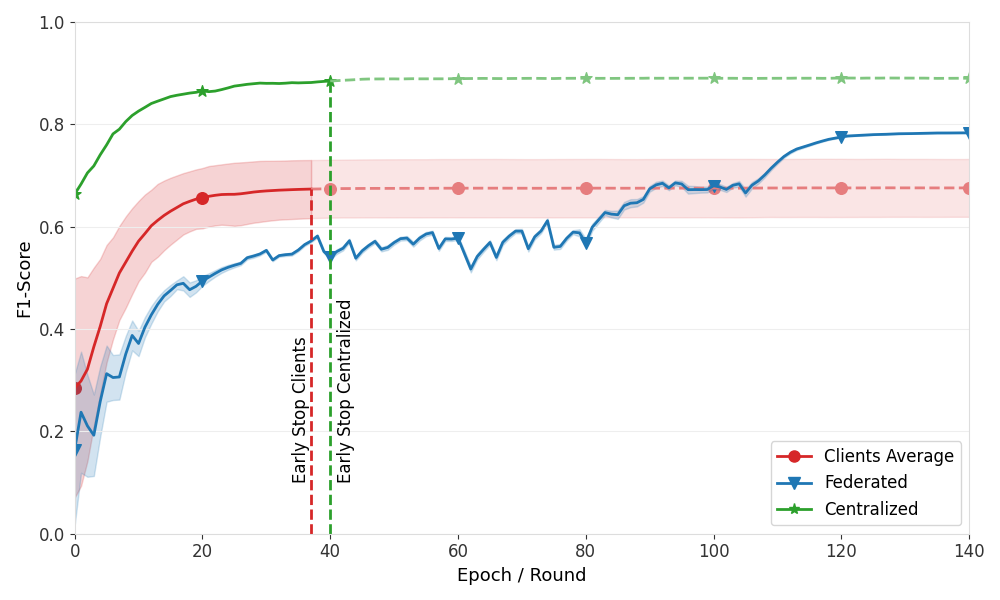}
    \vspace{-15pt}
     \caption{F1-Score Performance Comparison: training was extended beyond the early stopping for better representation, continuing until Federated Learning completed. Exponential moving average was applied to smooth the plot, and the confidence intervals reflect the standard deviation among the clients.}
     \label{fig:comparison}
     \vspace{-10pt}
\end{figure}

Finally, to evaluate the relative performance of the federated model, we compared it with local and centralized training. The best metrics for the three training configurations are shown in Table~\ref{tab:results4} and Figure~\ref{fig:comparison}. The average performance of local training across 8 simulated nodes was lower than that of the federated models, with an F1-score of 67\%. Centralized training achieved the highest overall performance with an F1-score of 89\%.


\begin{table}[t]
\caption{F1-Score/Accuracy comparison of normalization and aggregation algorithms.}
\vspace{-5pt}
\centering
\begin{tabularx}{0.78\columnwidth}{cccc}
& Batch Norm & Layer Norm & Group Norm \\ \hline
\multicolumn{1}{c|}{FedAvg} & \textbf{0.71 / 0.73} & 0.37 / 0.53 & 0.52 / 0.56 \\
\multicolumn{1}{c|}{FedDyn} & 0.41 / 0.42 & - & - \\
\multicolumn{1}{c|}{FedProx} & 0.44 / 0.49 & - & - \\
\multicolumn{1}{c|}{Scaffold} & \textbf{0.74 / 0.75} & 0.53 / 0.47 & 0.37 / 0.53 \\ \hline
\end{tabularx}
\label{tab:results1}
\vspace{-5pt}
\end{table}


\begin{table}[t]
\caption{F1-Score/Accuracy comparison of optimizers.}
\vspace{-5pt}
\centering
\begin{tabularx}{0.73\columnwidth}{cccc}
& Adam & SGD 0.001 & SGD 0.01 \\ \hline
\multicolumn{1}{c|}{FedAvg} & \textbf{0.71 / 0.73} & 0.50 / 0.58 & 0.40 / 0.57 \\
\multicolumn{1}{c|}{Scaffold} & \textbf{0.74 / 0.75} & 0.37 / 0.53 & 0.37 / 0.53 \\ \hline
\end{tabularx}
\label{tab:results2}
\vspace{-5pt}
\end{table}


\begin{table}[t]
\caption{F1-Score/Accuracy comparison of model sizes.}
\vspace{-10pt}
\centering
\begin{tabularx}{0.563\columnwidth}{ccc}
& Default size & Small size \\ \hline
\multicolumn{1}{c|}{FedAvg} & 0.71 / 0.73 & 0.62 / 0.66 \\
\multicolumn{1}{c|}{Scaffold} & 0.74 / 0.75 & \textbf{0.77 / 0.78} \\ \hline
\end{tabularx}
\label{tab:results3}
\vspace{-10pt}
\end{table}

\begin{table}[t]
\caption{F1-Score comparison of local, centralized and federated training.}
\vspace{-10pt}
\centering
\begin{tabularx}{0.48\columnwidth}{ccc}
Local & Centralized & Federated\\ \hline
0.67 &  0.89 & 0.77 \\
\end{tabularx}
\label{tab:results4}
\vspace{-10pt}
\end{table}

\section{Discussion}

In this study, we evaluated the performance of a machine learning model trained under a FL environment to detect atrial fibrillation.

Comparison of the best local, centralized, and federated models is crucial to understanding and interpreting the observed performance. The centralized scenario represents the optimal technical solution from an ML perspective, but is unfeasible in real-world healthcare applications due to data privacy concerns and regulations. As expected, this centralized scenario obtained the best result and is in line with SOTA \cite{Isaksen2022}. When each node trained only with its local data, a significant drop in performance was observed. The model trained with a limited amount of data was unable to generalize to the test set, overfitting to the local training data. These two performance metrics set crucial baselines for interpreting federated performance. The optimal federated configuration showed a notable improvement over the local training results, underscoring the benefits of FL. The federated approach managed to recover approximately half of the performance gap between centralized and local training. This research, which involves a federation of 8,000 ECGs spread across 8 nodes, suggests that federating hundreds of thousands of ECGs collected from EHR, IoT devices, and Holter ECGs could potentially yield even better performance, ultimately surpassing the centralized performance benchmark of the Physionet 2017 dataset.

Non-IID data is frequently mentioned as a challenge for federated machine learning models~\cite{LiFEDBN:NORMALIZATION}. Batch normalization layers, in particular, pose issues in federated settings due to varying means and standard deviations across the local nodes' datasets. Alternative aggregation methods to FedAvg aim to improve the performance of the federated model and address the challenges of different data distributions \cite{LiFEDBN:NORMALIZATION}. ECG devices and data collection have advanced over more than a century and are now fairly standard. Furthermore, all ECGs in the data set were collected using the same device, and the class distribution is relatively uniform between nodes, so we do not anticipate significant issues with the IID of the data. However, ECGs inherently vary from person to person, and as a personal device, AliveCor's collection methods also shows variability. Therefore, this study examines the impact of various aggregation and normalization techniques on network performance. Regarding aggregation algorithms, SCAFFOLD achieved the best results, while FedAvg also showed benefits over local training. FedDyn and FedProx exhibited performance issues and training was unsuccessful. Both algorithms use a proximal term correction that might be problematic for the type of network or the data used. Possibly due to the low non-IIDness of this dataset, the proximal term hindered the learning process rather than aiding it. The use of alternative normalization layers to address BN issues in FL is commonly proposed in the literature \cite{casella2023experimenting}. However, employing layer and group normalization led to a significant drop in model performance.

We also compared the use of different optmizers for the learning process. Adam was initially used because is the default optimizer for the network utilized; however, in FL scenarios, SGD is the optimizer most frequently adopted \cite{keskar2017improving}. Adam consistently outperformed SGD in all configurations tested by a significant margin. This indicates that the selection of an optimizer can greatly influence the performance of federated learning models.

As mentioned previously, we noticed significant overfitting to the local node data during the federated training of the global model. After several rounds, the global model achieved perfect performance on the local data within a few local batches but failed to generalize to the test dataset. Consequently, we reduced the network size from 32 convolutional blocks to 16 to mitigate overfitting to the local node data. This smaller neural network exhibited less overfitting to the training data, leading to improved test performance. The analysis of local training metrics proved to be essential for this interpretation and might not always be available in federated environments. Furthermore, it indicates that the optimal centralized architecture may not always be the best choice for federated learning.

\section{Conclusions and Future Work}

This study demonstrates the successful federation of neural networks for the early detection of AFib from raw ECG data.  SCAFFOLD and FedAvg were the most effective aggregation algorithms, outperforming local training. We have also examined the impact of normalization techniques and optimization methods on model performance. BN remained the most effective, while Adam optimizer outperformed SGD. The overfitting issues in the local node data were mitigated by reducing the network size, improving the generalizability of the model. Future studies should aim to incorporate more heterogeneous data, expanding the FL network, better diagnose and address overfitting problems, and address explainability in FL contexts. Addressing these areas will advance FL applications in healthcare, leading to more accurate, transparent and privacy-preserving diagnostic tools that enhance patient care and trust in AI-driven healthcare solutions.

\bibliographystyle{ieeetr}
\bibliography{refs}

\end{document}